\documentclass[%
nofootinbib,
twocolumn,
amsmath,amssymb,
aps,
]{revtex4-2}
\usepackage{graphicx}
\usepackage{bm}
\usepackage{longtable}
\usepackage{multirow}
\usepackage{braket}
\usepackage{xfrac} 
\usepackage{hyperref}
\usepackage{siunitx}
\DeclareMathAlphabet{\mathdefault}{OT1}{cmr}{m}{n}
\usepackage{pgf}
\usepackage{pgfplots}
\begin{document}

\preprint{}

\title[]{High-Throughput Exploration of NV-like Color Centers Across Host Materials}

\author{Oscar Groppfeldt}
\email{oscar.groppfeldt@liu.se}
\affiliation{Department of Physics, Chemistry and Biology, Linköping University, Linköping, Sweden}
\author{Joel Davidsson}
\email{joel.davidsson@liu.se}
\affiliation{Department of Physics, Chemistry and Biology, Linköping University, Linköping, Sweden}
\author{Rickard Armiento}
\email{rickard.armiento@liu.se}
\affiliation{Department of Physics, Chemistry and Biology, Linköping University, Linköping, Sweden}

\begin{abstract}
Point defects in semiconductors offer a promising platform for advancing quantum technologies due to their localized energy states and controllable spin properties. Prior research has focused on defects within a limited set of  materials such as diamond, silicon carbide, and hexagonal boron nitride. 
We present a high-throughput study to systematically identify and evaluate point defects across a diverse range of host materials, aiming to uncover previously unexplored defects in novel host materials suitable for use in quantum applications.
A range of host materials are selected for their desirable properties, such as appropriate bandgaps, crystal structures, and absence of d- or f-electrons. The Automatic Defect Analysis and Qualification (ADAQ) software framework is used to generate vacancies, substitutions with s- and p-elements, and interstitials in these materials and use density functional theory to calculate key properties such as Zero-Phonon Lines (ZPLs), ionic displacements, Transition Dipole Moments (TDMs), and formation energies. Special attention is given to charge correction methods for materials with dielectric anisotropy.
We uncover new defect-host combinations with advantageous properties for quantum applications: 13 defects across 29 host materials show properties similar to the nitrogen-vacancy (NV) center in diamond. Beryllium (Be) substitutional defects in SrS, MgS, and SrO emerge as particularly promising.
These findings contribute to diversifying and enhancing the materials available for quantum technologies.
\end{abstract} %%%%%%%%%

\maketitle

\section{Introduction}
\label{intro}
The potential of quantum technologies to revolutionize fields such as computing, cryptography, and sensing, drive the need for stable physical systems capable of supporting these advancements. Among various candidates, point defects in semiconductors with localized energy states due to atomic imperfections provide a promising platform for quantum applications~\cite{zhang_material_2020, awschalom_quantum_2018}.
Quantum applications require systems with specific physical characteristics, such as stable spin states and controllable optical transitions, to enable applications like quantum computing, communication, and sensing. Point defects in semiconductors have emerged as one such system, with defects like the nitrogen-vacancy (NV) center in diamond~\cite{davies1976, gali_ab_2019, doherty_nitrogen-vacancy_2013} already being fundamental components of various quantum devices and applications. However, as the number of applications of these technologies grows, so does the need for alternative host materials and defect configurations~\cite{wolfowicz_quantum_2021}. With this growing demand, it is important to understand the fundamental properties that make point defects viable for quantum applications. By examining the characteristics, we can identify new defect systems and configurations that meet the criteria of emerging technologies.

Weber \emph{et al.}\ lists five critical characteristics needed to ensure point defects are stable and efficient when participating in quantum operations as qubits~\cite{weber2010quantum}. They need to (\textit{i}) be long-lived bound states with spin sublevels; (\textit{ii}) allow optical pumping for state polarization; (\textit{iii}) have sublevel-dependent luminescence; (\textit{iv}) have interference-free optical transitions; and (\textit{v}) their bound states must have thermal stability. This list was derived from the properties observed for the nitrogen-vacancy (NV) center in diamond, the predominant defect for quantum applications.
The potential of systematic high-throughput density-functional theory (DFT) \cite{HK,KS} calculations to identify promising candidates for quantum applications have been demonstrated by recent studies on point defects in 2D and 3D materials. Highlights include, e.g., several studies on materials like diamond~\cite{luhmann2018screening, davidsson2024diamond}, silicon~\cite{xiong2023high, xiong2024discovery, ivanov2023database}, hexagonal boron nitride~\cite{tawfik2017first,vaidya2023quantum, cholsuk2024hbn}, and tungsten disulfide~\cite{thomas2024substitutional}. Extensive searches have also been done in wider ranges of 2D materials. For example, Bertoldo \textit{et al.}\ examined 1900 defects in 82 different 2D semiconductors and insulators~\cite{bertoldo2022quantum}. In a similar study, Huang \textit{et al.} produced a database with over 11,000 defect configurations across six 2D materials~\cite{huang2023unveiling}. These kinds of studies across wide ranges of host materials not only uncover defects with remarkable capabilities to characterize in more detail, but also provide quantitative insight into trends of defects in the materials. The authors are not aware of a database with defect data over a similar range of bulk host materials. 

In this work, we aim to address the lack of known point defects with interesting properties in less studied materials by a systematic high-throughput study across a wide range of host materials. We screen
all single point defects consisting of substitutional and interstitial s- and p-elements, as well as vacancies in 29 host materials selected from a list of suitable materials for hosting quantum defects compiled by Ferrenti \textit{et al.}~\cite{ferrenti2020identifying}. This list was narrowed down based on symmetry, band gaps, and stability (see Sec.~\ref{sec:host-materials}). The screening is performed using the Automatic Defect Analysis and Qualification (ADAQ) software framework~\cite{davidsson2021adaq} implemented using the high-throughput toolkit (\textit{httk})~\cite{armiento2020database}, to generate the defect systems and accurately calculate their magneto-optical properties. ADAQ has previously been successfully used to discover interesting defects in SiC~\cite{davidsson2022exhaustive}, CaO~\cite{davidsson2024discovery}, diamond~\cite{davidsson2024diamond}, and MgO~\cite{somjit2024nv}. 

\section{Defect properties}
The following subsections present essential properties of defects to determine their viability for quantum information applications. 

\subsection{Zero-phonon Line (ZPL)}
A key property for quantum applications is the ability of a defect to emit photons at specific wavelengths. This ability stems from the zero-phonon line (ZPL), which describes the photon emission from a relaxation from the excited state to the ground state where no phonon interactions occur. Without any interactions from surrounding phonons, the emitted photon carries the entire energy difference between the states, ensuring that it has the same wavelength every time it is emitted, which gives for the ZPL
\begin{equation}
    E_{\mathrm{ZPL}} = E_{e,\mathrm{min}} - E_{g,\mathrm{min}},
\end{equation}
where $E_{g,\mathrm{min}}$ and $E_{e,\mathrm{min}}$ are the energies of the ground and excited states.

\subsection{Huang-Rhys factor}
The Huang-Rhys factor $S$ quantifies the coupling between electronic states and vibrational modes~\cite{huang1950theory}. Defects with lower Huang-Rhys factors exhibit minimal vibrational losses, increasing the probability that an emission occurs through the ZPL. The Huang-Rhys factor is computationally expensive to calculate, but it can be estimated using a 1D model based on the one-phonon approximation, where
\begin{equation}
    \begin{split}
    S &= \sum_k S_k \text{, with}\\
    S_k &= \frac{\omega_k q_k^2}{2\hbar}\text{, and}\\
    q_k^2 &= \sum_i m_i \left|R_{e_i} - R_{g_i}\right|.
    \end{split}
\end{equation}
Here, $k$ indexes the phonon mode with frequency $\omega_k$ and $q_k$ is the sum of ionic displacements between the excited state $e_i$ and ground state $g_i$ over all ions $i$ with weight $m_i$~\cite{davidsson2021adaq, alkauskas2014first}. From the Huang-Rhys factor we calculate the Debye-Waller factor (DW), through the relation $\mathrm{DW} = e^{-S}$.

\subsection{Transition Dipole Moment (TDM)}
As an electron absorbs a photon, it transitions from its ground state $\ket{\psi_g}$ to an excited state $\ket{\psi_e}$. The probability of the specific transition $\ket{\psi_g} \rightarrow \ket{\psi_e}$ is given by the associated transition dipole moment $\mu_{ge}$. Using the Born-Oppenheimer approximation~\cite{oppenheimer1927born}, the TDM is given by
\begin{equation}
    \hat{\mu} = \bra{\psi_g}q\hat{r}\ket{\psi_e} = \frac{i\hbar}{(\varepsilon_{e} - \varepsilon_{g})m}\bra{\psi_e}\hat{p}\ket{\psi_g},
\end{equation}
where $\varepsilon_i$ is the energy of state $i$. The TDM is of particular interest as it directly relates to the intensity of the ZPL. The wavefunctions $\ket{\psi_e}$ and $\ket{\psi_g}$ are taken from the WAVECARs produced by VASP for the excited state and ground state calculations, respectively~\cite{davidsson2020theoretical}.

\subsection{Formation energy}
The energy required for a defect to form is given by its formation energy $\Delta H_{D,q}$. It is calculated as
\begin{equation}
    \Delta H_{D,q}(E_f,\mu) = (E_{D,q} - E_H) + \sum_in_i\mu_i + qE_f + E_{corr}(q)
\end{equation}
for a defect with charge $q$. Here, $E_{D,q} - E_H$ is the difference in energy between the system containing the defect and the host, $E_f$ is the Fermi energy, and the sum runs over all atomic species $i$ that have been removed ($n_i > 0$) or added ($n_i < 0$), scaled by their chemical potential $\mu_i$. Furthermore, $E_{corr}$ is a correction for the error in energy arising from the periodic boundary conditions used in the DFT calculations. As some of the materials in this study are dielectrically anisotropic, the commonly used charge-correction method by Lany and Zunger ~\cite{lany2008assessment} is not directly applicable due to the dielectric constant being a tensor. An alternative is to use the charge correction by Kumagai and Oba~\cite{kumagai2014electrostatics}, which alleviates the issues for the anisotropic host materials, but comes at the cost of only having a first-order correction, whereas the Lany-Zunger method is of third order. In this study we use the Lany-Zunger method, and reconcile the issues for anisotropic materials by using an effective dielectric constant in place of the tensor. See appendix \ref{sec:charge-correction-app} for an in-depth explanation.

\subsection{Defect hull}
The defect hull is spanned by the defects with the lowest formation energy in each stoichiometry as a function of the Fermi level. Defects on the defect hull are considered thermodynamically stable relative to other defects of the same stoichiometric composition. The defect hull has successfully been leveraged to predict new defects in 4H-SiC, which were identified experimentally~\cite{davidsson2022exhaustive}. 

\begin{figure}
    \begin{minipage}{\linewidth}
    \centering
    \includegraphics[width=\linewidth]{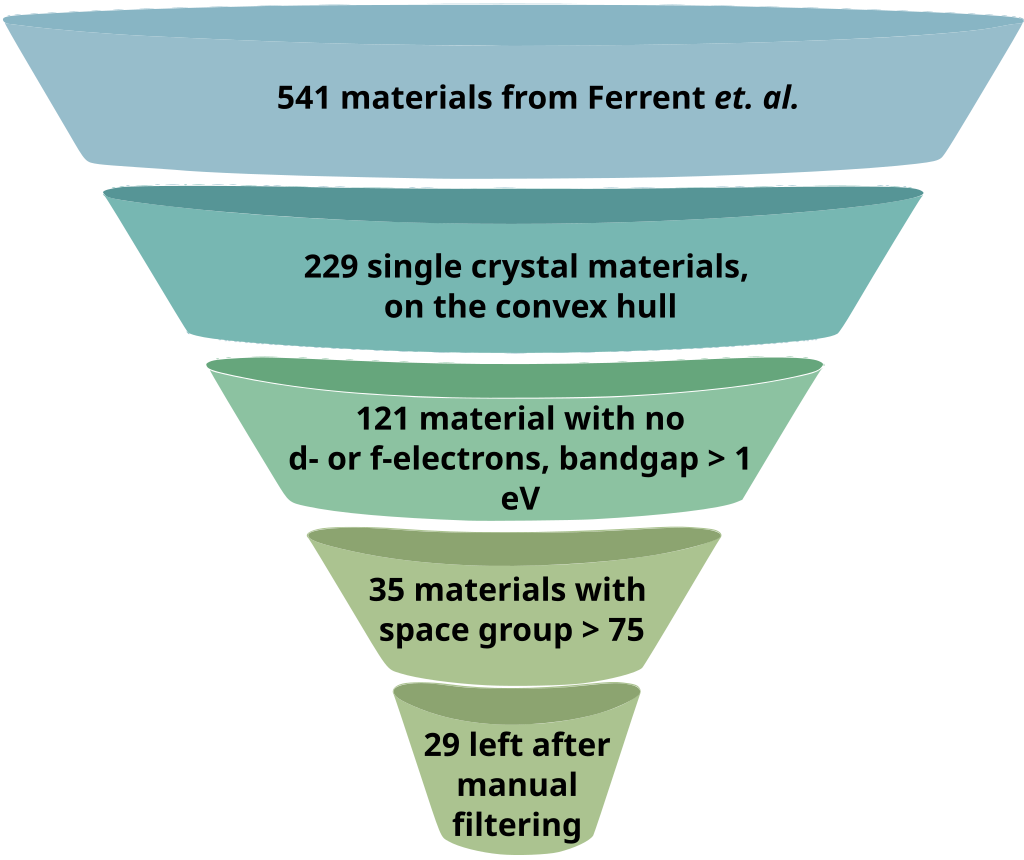}
    \caption{Overview of the filtering processes for the selection of host materials.}
    \label{fig:host-materials-funnel}
    \end{minipage}
\end{figure}

\section{Host material selection and defect generation}
\label{sec:host-materials}

The starting point for selecting the host materials to investigate in this work is a list of Ferrenti \textit{et al.}~\cite{ferrenti2020identifying} that identified 541 materials from the materials project database~\cite{jain2013commentary} as particularly suitable for hosting quantum defects.
From this list we select all (\textit{i}) single crystal materials (\textit{ii}) on the convex hull of thermodynamical stability (\textit{iii}) without d- or f-elements, (\textit{iv}) a band gap of $1.0$ eV or more. Furthermore, we also (\textit{vi}) remove materials with a space group number below 75 (discussed in detail in next paragraph), and (\textit{v}) have not previously investigated using ADAQ. The remaining materials are sorted by their quantum coherence times as calculated in Ref.~\onlinecite{kanai2022generalized}. As a final step we manually removed all materials containing lead ($\mathrm{Ba_2PbO_4}$, $\mathrm{Si_2Pb_3O7}$, $\mathrm{Pb_3O_4}$) and $\mathrm{CS_{14}}$, as $\mathrm{CS_{14}}$ did not appear to be a covalently bonded solid. The crystal structures are obtained using the Materials Project API~\cite{jain2013commentary}. Details on the resulting host materials are presented in Table \ref{tab:materials}.

We restrict our screening to host materials whose crystallographic point group admits \textit{essential} (symmetry-enforced) orbital degeneracies, i.e., E (or T) irreducible representations. Such protected triplets (or quartets) guarantee electronic states that cannot be split by symmetric perturbations, enhancing the resilience to local distortions. This stability is not present in defects with accidental degeneracies. Because the symmetry of the defect is limited by the symmetry of the host material, we select hosts whose crystallographic point groups contain E- and T-type irreducible representations~\cite{bassett2019quantum}. This excludes all non-complex abelian systems, i.e., the triclinic, monoclinic, and orthorombic crystal families (space group 1-74). We are therefore limited to tetragonal, trigonal, or hexagonal host materials for defects with spin 1, and cubic systems for defects with spin 1 or spin 3/2. This group-theoretic criteria underlies the success of defects such as $\mathrm{NV^-}$ in diamond ($\mathrm{C_{3v}}$) and divacancies in SiC ($\mathrm{C_{3v}}$)~\cite{dreyer_first-principles_2018}. 

\begin{table}
    \centering
    \begin{tabular}{llcS[table-format=2.3]}\label{tab:hosts}
        Host & mp-id & Space group & \multicolumn{1}{c}{$T_2$ (ms)~\cite{kanai2022generalized}} \\
        \hline
        $\mathrm{CaS}$ & mp-1672 & 225 & 22.502 \\
        $\mathrm{CaCO_{3}}$ & mp-3953 & 167 & 10.524 \\
        $\mathrm{CaMg_2(SO_4)_3}$ & mp-554094 & 176 & 3.720 \\
        $\mathrm{BaGe_{2}S_{5}}$ & mp-28710 & 227 & 3.137 \\
        $\mathrm{Sr_{2}MgGe_{2}O_{7}}$ & mp-972387 & 113 & 2.761 \\
        $\mathrm{Sr_{2}MgSi_{2}O_{7}}$ & mp-6564 & 113 & 2.600 \\
        $\mathrm{Ca_{2}Ge_{7}O_{16}}$ & mp-29273 & 117 & 2.478 \\
        $\mathrm{SrGe_{4}O_{9}}$ & mp-9380 & 150 & 2.446 \\
        $\mathrm{SrS}$ & mp-1087 & 225 & 2.387 \\
        $\mathrm{BaGe_{4}O_{9}}$ & mp-3848 & 150 & 2.073 \\
        $\mathrm{CaSe}$ & mp-1415 & 225 & 1.985 \\
        $\mathrm{SeO_2}$ & mp-726 & 135 & 1.847 \\
        $\mathrm{SrSe}$ & mp-2758 & 225 & 1.767 \\
        $\mathrm{BaS_{3}}$ & mp-239 & 113 & 1.722 \\
        $\mathrm{Ba_{2}MgGe_{2}O_{7}}$ & mp-1190545 & 113 & 1.583 \\
        $\mathrm{GeSe_{2}}$ & mp-10074 & 122 & 1.577 \\
        $\mathrm{MgCO_{3}}$ & mp-5348 & 167 & 1.537 \\
        $\mathrm{SrO}$ & mp-2472 & 225 & 1.516 \\
        $\mathrm{Ba_{2}MgSi_{2}O_{7}}$ & mp-9338 & 113 & 1.478 \\
        $\mathrm{SrTe}$ & mp-1958 & 225 & 1.227 \\
        $\mathrm{MgS}$ & mp-1315 & 225 & 1.156 \\
        $\mathrm{BaSe}$ & mp-1253 & 225 & 1.143 \\
        $\mathrm{BaS}$ & mp-1500 & 225 & 1.127 \\
        $\mathrm{Sn(SeO_3)_2}$ & mp-556672 & 205 & 1.123 \\
        $\mathrm{BaTe}$ & mp-1000 & 225 & 1.076 \\
        $\mathrm{Mg_{3}TeO_{6}}$ & mp-3118 & 148 & 1.050 \\
        $\mathrm{Ba_{2}TeO}$ & mp-1078191 & 129 & 1.007 \\
        $\mathrm{BaO_{2}}$ & mp-1105 & 139 & 0.874 \\
        $\mathrm{BaO}$ & mp-1342 & 225 & 0.742 \\
    \end{tabular}
    \caption{The selected host materials along with their identifier in the materials project database. The calculated $T_2$ times are taken from Ref.~\cite{kanai2022generalized}.}
    \label{tab:materials}
\end{table}

\begin{figure*}
\begin{minipage}{\linewidth}
\begin{longtable*}{
    p{7em} % Host name
    p{7em} % Defect type
    S[table-format=1.2] % Charge
    c % Spin
    S[table-format=1.2]  % ZPL
    S[table-format=2.2]  % TDM
    S[table-format=1.2]  % deltaQ
    S[table-format=2.2]  % DW
    }
    \caption{Defect systems on the defect hull with a converged ZPL $>$ 0.5 eV, spin $\geq$ 1, a TDM larger than 3 Debye, and an estimated Debye-Waller (DW) factor $>$ 1\%.}
    \label{tab:iso-defects}\\
    \textbf{Host} 
        & \textbf{Defect type} 
        & \textbf{Charge} 
        & \textbf{Spin} 
        & \textbf{ZPL (eV)}
        & \multicolumn{1}{c}{\textbf{TDM (Debye)}}
        & \boldsymbol{$\Delta Q$} \textbf{(amu$^{1/2}$)}
        & \multicolumn{1}{c}{\textbf{DW (\%)}}
    \\
  \endfirsthead
    \hline
    \textbf{Host} 
        & \textbf{Defect type} 
        & \textbf{Charge} 
        & \textbf{Spin} 
        & \textbf{ZPL (eV)}
        & \multicolumn{1}{c}{\textbf{TDM (Debye)}}
        & \boldsymbol{$\Delta Q$} \textbf{(amu$^{1/2}$)}
        & \multicolumn{1}{c}{\textbf{DW (\%)}}
    \\
    \endhead
    \endfoot
    \endlastfoot
    \hline
    \multirow{1}{*}{MgS} & $\mathrm{Be_S}$ & 0 & 1 & 1.94 & 9.22 & 0.65 & 21.15\\
    \hline
    \multirow{2}{*}{$\mathrm{Ca_2Ge_7O_{16}}$} & $\mathrm{F_{Ca}}$ & 0 & $\frac{3}{2}$ & 0.55 & 5.37 & 0.54 & 20.19\\
    & $\mathrm{Rb_{Ge}}$ & 0 & $\frac{3}{2}$ & 0.70 & 6.40 & 0.84 & 7.45\\
    \hline
    \multirow{1}{*}{$\mathrm{Mg_3TeO_6}$} & $\mathrm{Ba_{Te}}$ & 0 & 2 & 0.76 & 9.95 & 0.94 & 1.98\\
    \hline
    \multirow{2}{*}{SrO} & $\mathrm{Int_{Sb}}$ & 1 & 1 & 1.62 & 5.44 & 0.69 & 14.12\\
    & $\mathrm{Be_O}$ & 0 & 1 & 1.95 & 8.51 & 0.67 & 25.79\\
    \hline
    \multirow{1}{*}{SrS} & $\mathrm{Be_S}$ & 0 & 1 & 2.06 & 8.22 & 0.55 & 44.39\\
    \hline
    \multirow{1}{*}{$\mathrm{Sr_2MgSi_2O_7}$} & $\mathrm{K_{Si}}$ & 0 & $\frac{3}{2}$ & 0.53 & 3.92 & 1.23 & 2.55 \\
    \hline
    \multirow{4}{*}{$\mathrm{Sr_2MgGe_2O_7}$} & $\mathrm{K_{Ge}}$ & -1 & 1 & 0.51 & 25.85 & 1.16 & 1.38\\
    & $\mathrm{F_{Ge}}$ & 0 & $\frac{3}{2}$ & 0.51 & 4.14 & 1.26 & 1.33\\
    & $\mathrm{F_{Ge}}$ & -1 & 1 & 0.55 & 8.98 & 0.75 & 14.53\\
    & $\mathrm{Rb_{Ge}}$ & -1 & 1 & 0.54 & 5.30 & 0.53 & 29.64\\
    \hline
    \multirow{1}{*}{$\mathrm{Ba_2MgSi_2O_7}$} & $\mathrm{Ba_{Si}}$ & 0 & 1 & 0.52 & 4.33 & 1.21 & 2.19\\
    \hline
\end{longtable*}
\end{minipage}
\end{figure*}

For each host, we generate a set of single point defects, including vacancies, substitutionals, and interstitials. The generation is limited by a number of constraints: (\textit{i}) \emph{Spatial constraints:} the minimum lattice parameter length of the supercell is taken to be 20 Å to ensure that the generated defects are well separated spatially and single-point defects are generated individually, with no clusters or extended defects;\footnote{This constraint was lifted to 10 Å for the $\mathrm{SrGe_4O_9}$, $\mathrm{BaGe_4O_9}$, and $\mathrm{BaGe_2S_5}$ host systems due to the size of the unitcell and their computational complexity.} (\textit{ii}) \emph{Substitutional constrains:} dopand species are only allowed to be intrinsic or elements from the s- and p-blocks of the periodic table; (\textit{iii}) \emph{Interstitial placement constraints:} distances between interstitial atoms and surrounding host atoms are restricted to a range of 1 Å to 3.5 Å.

\section{Computational details}
We use the ADAQ framework~\cite{davidsson2021adaq} implemented using the high-throughput toolkit (\emph{httk})~\cite{armiento2020database} to generate defects, run automated computational workflows, calculate relevant properties, and identify defects of interest. The calculations uses Kohn-Sham DFT \cite{HK,KS} implemented in the Vienna Ab initio Simulation Package (VASP)~\cite{kresse1994ab} with the exchange-correlation functional by Perdew, Burke and Ernzerhof (PBE)~\cite{perdew1996generalized}. The convergence settings for screening-level ADAQ calculations from Ref.~\cite{davidsson2021adaq} are used. These convergence settings include a plane wave energy cutoff of 600 eV and a kinetic energy cutoff of 900 eV, which are chosen to cover the requirements of a wide range of elements. In addition, the Fast Fourier Transform (FFT) grid is set to twice the largest wave vector, and a $\Gamma$-centered Monkhost-Pack k-point grid is used. When performing calculations at the $\Gamma$-point only, Fermi smearing with a width of 1 meV is appliced. No symmetry is applied for the defect calculations. 

\section{Results}
In the following sections, we first examine how the high-spin states are distributed across different host system symmetries, and then focus on the identification and characterization of NV-like defects.

\subsection{Spin data}
\begin{figure*}[t]
    \centering
    \includegraphics[width=\textwidth]{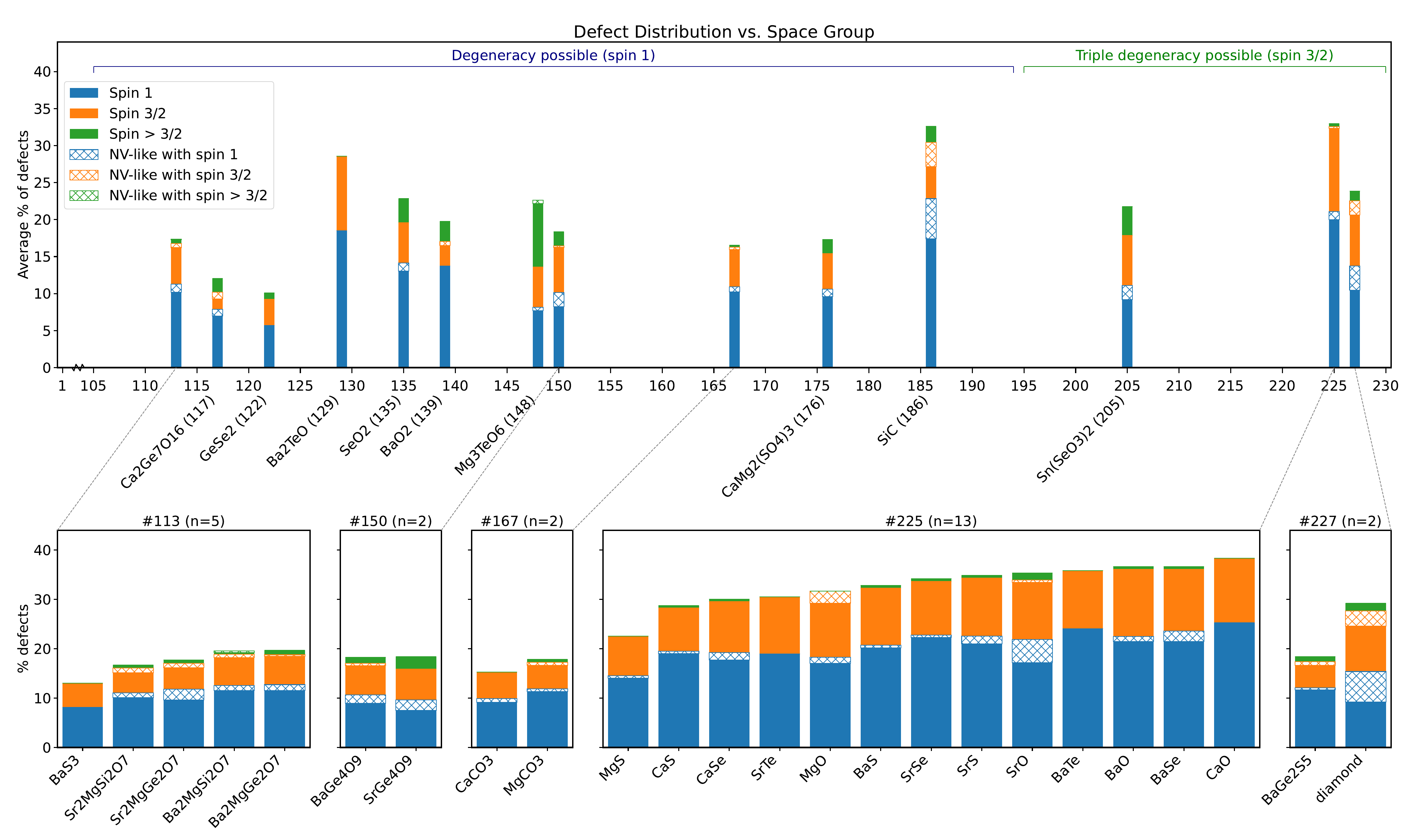}
    \caption{Average percentage of point defect by spin state across the space groups of the host material. The main panel shows, for each space group, the mean fraction of defects with spin 1 (solid blue), spin 3/2 (solid orange), and spin $>$ 3/2 (solid green); hashed bars indicate the subset of NV-like centers. Insets display the individual material data for spacegroup with multiple host materials.}
  \label{fig:spin-distribution}
\end{figure*}

As a large number of host materials are evaluated in this study, we compare the number of stable defects that have high spin between different host symmetries. The results of this comparison is found in Figure \ref{fig:spin-distribution}, where each bar represents the fraction of stable defects that exhibit high spin compared to all stable defects. The figure is divided into subpanels for each space group of the (perfect) host crystal. For space groups shared between multiple host materials, the main panel shows the arithmetic average of all hosts in that space group. Single defect data for MgO, CaO, 4H-SiC, and diamond (materials previously studied using ADAQ~\cite{somjit2024nv, davidsson2024discovery,davidsson2021adaq, davidsson2024diamond}) are also added for comparison.

\subsection{NV-like defects}
As the next step, we investigate point defects in host materials for key properties such as ZPL energies, TDM values, and defect stability. Table~\ref{tab:materials} shows the host materials investigated and their Materials Project IDs. A key highlight is the beryllium substitutionals, which appears in three different host materials MgS, SrS, and SrO. The formation energies for these substitutionals are shown in Figure \ref{fig:formation-SrS_Be}, \ref{fig:formation-MgS_Be}, and \ref{fig:formation-SrO_Be}, respectively.

\begin{figure}%
    \centering
    \resizebox{0.5\textwidth}{!}{\input{SrS.pgf}}
    \caption{Formation energy for $\mathrm{Be_S}$ in SrS for different charges $q$ (indicated by color) and spins (indicated by line dashing).}
    \label{fig:formation-SrS_Be}
\end{figure}%
\begin{figure}%
    \centering
    \resizebox{0.5\textwidth}{!}{\input{MgS.pgf}}
    \caption{Formation energy for $\mathrm{Be_S}$ in MgS for different charges $q$ (indicated by color) and spins (indicated by line dashing).}
    \label{fig:formation-MgS_Be}
\end{figure}%
\begin{figure}%
    \centering
    \resizebox{0.5\textwidth}{!}{\input{SrO.pgf}}
    \caption{Formation energy for $\mathrm{Be_O}$ in SrO for different charges $q$ (indicated by color) and spins (indicated by line dashing).}
    \label{fig:formation-SrO_Be}
\end{figure}%

\section{Discussion}
%\subsection{Spin data}
Based on the discussion in Sec.~\ref{sec:host-materials} regarding how the space group of the host material affects the spin of defects, one would expect that a pattern should emerge in Figure~\ref{fig:spin-distribution}, with hosts in higher space groups having more defects with high spin due to the higher dimensional irreducible representations. However, such a pattern is not easily discernible from the data in this study. Hence, our results suggest a significant contribution of high-spin defects from accidental degeneracies. These degeneracies are possible when the exchange interaction between electrons in different defect states exceed the single-particle energy gap separating those states~\cite{bassett2019quantum}. The unexpected abundance of spin $3/2$ defects in $\mathrm{Ba_2TeO}$ and spin $\geq 3/2$ defects in $\mathrm{Mg_3TeO_6}$ stand out among the other host materials. $\mathrm{Ba_2TeO}$ is a layered material where sheets derived from two cubic parent materials are stacked~\cite{besara2015ba2teo}. In this case, the two parent host materials appear to create local pseudo-symmetries that are cubic in nature, allowing for triple degeneracies, despite the host material as a whole lacking this symmetry. In $\mathrm{Mg_3TeO_6}$, both Mg and Te are octahedrally coordinated~\cite{newnham1970crystal}, which provides a local $\mathrm{O_h}$ pseudo-symmetry. Indeed, defects on the Te site alone account for all but one defect with spin $>1$ in our data, the remaining defect is a substitution on a Mg site.
It is a central result from this work that these local pseudo-symmetries appear as a key mechanism for high-spin defects relevant for quantum applications via approximate accidental degeneracies. 

Strontium oxide (SrO) stands out compared to the other materials in this study by hosting as many as 11 high-spin NV-like defects. This prevalence of NV-like defects, about 5\% of the ones considered in this work, approaches that of diamond and 4H-SiC. SrO is a rock-salt oxide, similar to CaO and MgO recently suggested as promising in Ref.~\cite{davidsson2024discovery}. 
SrO has a large static dielectric constant ($\epsilon \approx 16$) ~\cite{jacobson1968infrared} which can be important for deep charged point defects, where the screening can isolate them from the surrounding electronic states. The most common isotopes, $\mathrm{^{88}Sr}$ (approximately 83\% abundance) and $\mathrm{^{16}O}$ (over 99\% abundance), have zero nuclear spin, which means the nuclear spin bath in a typical (non-purified) SrO lattice is minimal. These two attributes, along with the high $T_2$ time of SrO (1.5 ms, ~\cite{kanai2022generalized}) makes it a promising candidate for realizing deep, stable spin centers.

%\subsection{NV-like defects}
One of the NV-like defects in SrO is a Beryllium substitutional. Across all materials, beryllium occurs three times as a substitutional defect, all with similar properties. Their $\Delta Q$ range from $0.55\ \mathrm{amu}^{1/2}$ to $0.67\ \mathrm{amu}^{1/2}$, which indicates that a large portion of their emissions occur in the ZPL. The ZPLs also have a comparatively strong intensity, with the TDMs being between $8.22\ \mathrm{Debye}$ and $9.22\ \mathrm{Debye}$, with emissions estimated to be around $2\ \mathrm{eV}$, placing them in the red part of the visible spectrum. While these wavelengths are too short for direct use in current telecom optics,~\cite{ITU_G694.1_2020, ITU_G694.2_2020} they are close to that of the emissions of the $\mathrm{NV^-}$ center~\cite{alkauskas2014first}. This similarity is advantageous, as it is may be possible to adopt methods for adapting the emissions of the $\mathrm{NV^-}$ center to interoperate with telecom wavelengths, which have already been successfully demonstrated to work~\cite{dreau2018quantum}. While these beryllium substitutionals seem promising, note that the reported ZPL, TDM, and $\Delta Q$ values are only calculated to screening-level accuracy. Further verification by more accurate methods, such as DFT with the exchange-correlation functional of Heyd, Scuseria, and Ernzerhof (HSE06)~\cite{heyd2003hybrid}, should be performed to validate their viability.
We have only considered single point defects. In practice, defect clusters are commonly observed~\cite{wesch1989defect}. In particular, vacancies are frequently sufficiently mobile to be able to migrate into more energetically favorable positions where they can combine with other defects to form larger clusters~\cite{horiki1998identification, olsen2016investigation}. Hence, a natural future extension of this high-throughput study is to consider more complex defects, in particular vacancy complexes and defect-vacancy clusters; in particular in configurations with the right conditions for local pseudo-symmetry to allow approximate accidental degeneracy as discussed above.

\section{Conclusion}
In this work, we have provided a systematic exploration of point defects for quantum technologies across 29 host materials. By leveraging high-throughput computational methods, using DFT and the ADAQ framework, several new and promising defect systems have been identified. From our results, we have also uncovered that alongside the high-spin defects enabled by the degeneracies from essential symmetries for the defect in the host material, there is also a significant contribution from accidental approximate degeneracies enabled by the local symmetry around the defect.

Notably, beryllium substitutionals in SrS, MgS, and SrO emerge as particularly interesting due to their strong ZPL properties comparable to that of the well studied NV-center in diamond, and potentially offer improved photon emission characteristics.
This study opens for future works with a focus on refining the understanding of these systems through more computationally intensive techniques and experimental validation. By advancing the screening for, and characterization of, such defects, this study contributes to the ongoing effort to diversify and enhance the material platforms available for quantum technologies.
Ultimately, our high-throughput DFT screening reveals that promising NV-like defects — most notably the Be substitutions — exist outside of the commonly studied host materials. It also shows that SrO in particular offers an exceptionally favorable host environment for point defects aimed at quantum technology. These results both broaden the range of viable materials for quantum devices and set the stage for targeted hybrid-functional studies and experimental validation.

\section*{Acknowledgments}
We acknowledge support from the Swedish Research Council (VR) Grant No. 2022-00276, 2020-05402, and 2023-05358. 
The computations were enabled by resources provided by the National Academic Infrastructure for Supercomputing in Sweden (NAISS), partially funded by the Swedish Research Council through grant agreement no. 2022-06725.

\section*{Author contributions}
O.G. performed the calculations, made the figures, and wrote the original draft. J.D. conceptualized the project in discussion with R.A, supervised the computational methodology, contributed to data interpretation, reviewed, and edited the manuscript. R.A. supervised the project, provided methodological guidance, reviewed, and edited the manuscript.

\appendix
\section{Treatment of charge correction}\label{sec:charge-correction-app}
\begin{figure*}
    \centering
    \resizebox{\textwidth}{!}{\input{chargecorrection_scatter.pgf}}
    \caption{First order Makov-Payne charge correction as a function of Kumagai-Oba charge correction energy, using the \textit{isotropized} $\varepsilon$ (blue) and $\varepsilon_{\mathrm{eff}}$ from the method of Shang \textit{et al.}~\cite{shang_revisiting_2025}.}
    \label{fig:cce}
\end{figure*}
Accurate image-charge corrections are essential for reliable defect energetics in periodic supercell calculations. The classical Makov-Payne (MP) scheme~\cite{makov_periodic_1995} (and by extension the Lany-Zunger (LZ) formalism~\cite{lany2008assessment}) assumes a cubic supercell with an isotropic dielectric constant. Both MP and the Kumagai-Oba (KO) approach~\cite{kumagai2014electrostatics} are strictly \textit{first-order} in the inverse cell length $L^{-1}$, accounting only for the monopole image-charge term. The LZ correction goes one step further by incorporating higher-order multipole contributions (up to $L^{-3}$), thereby capturing \textit{third-order} effects within its formal first order correction~\cite{lany2008assessment}. However, neither MP nor LZ treats dielectric \textit{anisotropy}, whereas KO correctly handles the \textit{full} dielectric tensor in arbitrarily shaped supercells~\cite{kumagai2014electrostatics}.
In this work, we survey a wide range of host materials, only a subset of which strictly satisfy the isotropic, cubic assumptions of MP/LZ, to determine when explicit tensor treatment is essential. For each defect supercell we compute the monopole correction in three ways: (\emph{i}) MP using the arithmetic average of the dielectric-tensors eigenvalues; (\emph{ii}) MP using the effective dielectric consatant $\varepsilon_{\mathrm{eff}}$ of Shang \textit{et al.}~\cite{shang_revisiting_2025}; and (\emph{iii}) KO full-tensor first-order correction. We take the KO results as the basis of comparison, i.e., the ground truth, and apply all three schemes to the 190 materials (out of the 541 from Ferrenti \textit{et al.}~\cite{ferrenti2020identifying}) for which full dielectric tensors are available in the Materials Project. Figure \ref{fig:cce} plots the monopole energies from each LZ variant against KO. The energies are computed for supercell geometries of the host material. The supercells are constructed by taking the conventional cell of the material, and scaling each lattice vector $a_i$ of the conventional cell by a factor $f_i = \lfloor ||\bm{a_i}|| + \frac{1}{2} \rfloor$. We perform this reshaping to mirror what ADAQ does when creating the supercell used for the other defect calculations. On average both LZ approximations stay within approximately 120 meV of KO, largely because our large, nearly cubic, supercells both diminish the magnitude of the first-order term and reduce the impact of anisotropy. They perform well for materials with high anisotropy, such as $\mathrm{HfS_3}$, $\mathrm{PbSeO_3}$, and $\mathrm{Ba_2ZrS_4}$, but struggle when the anisotropy and/or the cubic nature of the material becomes too large. This is the case for $\mathrm{HfS_2}$, $\mathrm{PdSe_2O_5}$, $\mathrm{PdSeO_3}$, and $\mathrm{SnGeS_3}$.
The materials used for this comparison originate from the selection of host materials suitable for hosting quantum defects, and the results of this comparison should therefore not be taken as a general result beyond this context. However, we find that these results are sufficiently convincing to treat the supercells of non-cubic and dielectrically anisotropic host materials in this study as approximately fulfilling the formal requirements of using the LZ charge correction when using an isotropic approximation of the dielectric tensor. We therefore use the approximation of Shang \textit{et al.}~\cite{shang_revisiting_2025} to give us a dielectric constant to use with the LZ charge correction, in order to maintain consistency with previous entries in the ADAQ database.
\clearpage
\bibliography{ref}

\end{document}